\pdfoutput=1
\documentclass[prx,twocolumn,showpacs,amsmath,amssymb,floatfix,superscriptaddress]{revtex4-1}

\usepackage{graphicx}  
\usepackage{bm}    
\usepackage{braket}
\usepackage{appendix}
\usepackage[colorlinks=true, linkcolor=blue, urlcolor=blue,
citecolor=blue]{hyperref}

\newcommand{\RNum}[1]{\uppercase\expandafter{\romannumeral #1\relax}}

\begin{document}

\title{Magnon-driven chiral charge and spin pumping and electron-magnon scattering from time-dependent quantum transport combined with atomistic spin dynamics theory}
\author{Abhin Suresh}
\affiliation{Department of Physics and Astronomy, University of Delaware, Newark, DE 19716, USA}
\author{Utkarsh Bajpai}
\affiliation{Department of Physics and Astronomy, University of Delaware, Newark, DE 19716, USA}
\author{Branislav K. Nikoli\'{c}}
\email{bnikolic@udel.edu}
\affiliation{Department of Physics and Astronomy, University of Delaware, Newark, DE 19716, USA}
\affiliation{Kavli Institute for Theoretical Physics, University of California, Santa Barbara, CA
93106-4030}

\begin{abstract}
    Using newly developed quantum-classical hybrid framework, we investigate interaction between spin-polarized conduction electrons and a single spin wave (SW) coherently excited within a metallic ferromagnetic nanowire. The SW is described by classical atomistic spin dynamics as a collection of localized magnetic moments on each atom, which  precess as classical vectors with harmonic variation  in  the  phase of precession of adjacent  moments around the  local   magnetization  direction.  The conduction electrons are described quantum-mechanically using time-dependent nonequilibrium Green functions.  When the nanowire hosting SW is attached to two normal metal (NM) leads, with no dc bias voltage applied between them, the SW pumps {\em chiral} electronic charge and spin currents  into the leads---their direction is tied to the direction of SW propagation and they scale linearly with the frequency of the precession. This is in contrast to: standard pumping by the uniform precession mode with identical spin currents flowing in both directions and {\em no} accompanying charge current; or experimentally observed [C.~Ciccarelli {\em et al.},  Nat. Nanotech. \textbf{10}, 50 (2014); M.~Evelt {\em et al.}, Phys. Rev. B {\bf 95}, 024408 (2017)] magnonic charge pumping which {\em requires} spin-orbit coupling (SOC). The mechanism behind our prediction is {\em nonadiabaticity due to time-retardation effects}---motion of  localized magnetic moment affects conduction electron spin in a retarded way, so that it takes a finite time until the electron spin reacts to the motion of the classical vector. This makes the two vectors misaligned, even for {\em zero} SOC, and we visualize retardation effects by computing the spatial profile of nonadiabaticity angle between these two vectors across the nanowire. Upon injecting dc spin-polarized charge current from the left NM lead, electrons interact with SW where  outflowing charge and spin current into the right NM lead are changed due to both scattering off time-dependent potential generated by the SW and superposition with the currents pumped by the SW itself. Using Lorentzian voltage pulse to excite leviton out of the Fermi sea, which carries one electron charge with no accompanying electron-hole pairs and behaves as soliton-like quasiparticle, we describe how a single electron interacts with a single SW. 
\end{abstract}

%{\bf In this proposal, we present a novel view on magnons: the nonequilibrium
%	magnons, whether generated by the electric current via electron-magnon
%	scattering or by a thermal gradient, are actually efficient angular momentum carriers that can
%	be used for transferring magnetization information.}

\maketitle

\section{Introduction}

\begin{figure}[h!]
	\includegraphics[scale=0.9]{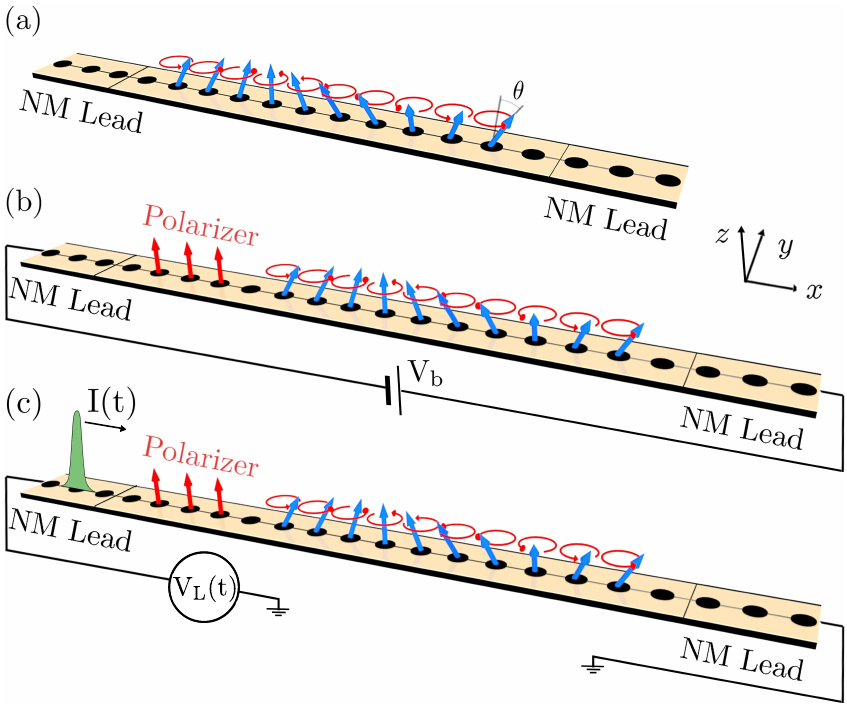}
	\caption{Schematic view of two-terminal setups where FM wire, modeled as 1D chain of ferromagnetic atoms~\cite{Spinelli2014}, hosts SW comprised of $N=10$  localized magnetic moments $\mathbf{M}_i(t)$ precessing as classical vectors with frequency $\omega$ and cone angle $\theta=10^\circ$~\cite{Demidov2007}, as well as with harmonic variation  in  the  phase  of precession of adjacent  moments. The wire is attached to L and R semi-infinite NM leads which terminate into the macroscopic reservoirs where in: (a) no bias voltage is applied between the reservoirs; (b) small bias voltage $V_b$ is applied to inject dc unpolarized charge current into the wire, which is then spin-polarized by three fixed spins (red arrows); and (c) Lorentzian voltage pulse~\cite{Ivanov1997,Keeling2006} is applied between the reservoirs to inject leviton current pulse $I_\mathrm{L}(t)$ into the wire carrying charge $\int\!\!dt\, I_\mathrm{L}(t) = 2e$.}
	\label{fig:fig1}
\end{figure}

In the semiclassical picture~\cite{Kim2010,Evans2014}, a spin wave (SW) is a  disturbance in the local magnetic ordering of a ferromagnetic material in which localized  magnetic moments precess around the easy axis with the phase of precession of adjacent moments varying harmonically in space over the  wavelength $\lambda$, as illustrated in Fig.~\ref{fig:fig1}. The quanta of energy of SW behave as a quasiparticle, termed magnon, which carries energy $\hbar \omega$ and spin $\hbar$. The frequency $\omega$ of the precession is commonly in GHz range of microwaves, but it can reach THz range in antiferromagnets~\cite{Jungfleisch2018}. The SWs can be excited in equilibrium as incoherent thermal fluctuations, which then reduce the total magnetization with increasing temperature~\cite{Hofmann2011}. They can also be excited by external fields~\cite{Zilberman1995,Demidov2007,Woo2017,Sandweg2011,Tzschaschel2017,Evelt2017} which leads to coherent propagation of SWs as a dispersive signal.  

Out of equilibrium, electron-magnon interaction is encountered in numerous  phenomena in spintronic devices, such as inside magnetic layers or at their interfaces with  layers of normal metals and insulators. For example, such processes can: increase resistivity of ferromagnetic metal (FM) with temperature due to spin-flip scattering from thermal spin disorder~\cite{Raquet2002,Starikov2018}; play an essential role in the laser-induced ultrafast demagnetization~\cite{Carpene2008}; generate nontrivial temperature and bias voltage dependence of tunneling magnetoresistance in magnetic tunnel junctions~\cite{Drewello2008,Mahfouzi2014}; open inelastic conducting channels~\cite{Slonczewski2007}; contribute to spin-transfer~\cite{Balashov2008,Cheng2019,Wang2019} and spin-orbit torques~\cite{Okuma2017}; and convert magnonic spin currents into electronic spin current or vice versa at magnetic-insulator/normal-metal interfaces~\cite{Kajiwara2010,Sandweg2011,Zheng2017,Uchida2010}. Magnon driven {\em chiral} charge pumping---where magnon generates electronic charge current in the absence of any bias voltage with the direction of current changing upon changing the direction of magnon propagation---has also been observed experimentally~\cite{Evelt2017,Ciccarelli2014}, while crucially relying on the presence of spin-orbit coupling (SOC).

The nonequilibrium many-body perturbation theory~\cite{Mahfouzi2014,Zheng2017}, formulated using Feynman diagrams for nonequilibrium Green function  (GF)~\cite{Stefanucci2013}, offers rigorous quantum-mechanical treatment of both electrons and magnons, once the original spin operators are mapped to the bosonic ones~\cite{Kiselev2000}. However, to ensure current conservation, one has to sum large classes of such diagrams~\cite{Mera2016} which can lead to errors due to missed vertex corrections~\cite{Gukelberger2015}. Furthermore, due to small magnon bandwidth, small electron-magnon interaction constant $J_{sd}$ in the realm of electrons can become strongly correlated regime for magnons due to large ratio \mbox{$J_{sd}/$magnon-bandwidth}. This can lead to quasibound states of magnons surrounded by electron-hole pairs~\cite{Mahfouzi2014}, therefore suggesting that complicated higher order diagrams should be evaluated. This severely limits system size in two-terminal geometries of Fig.~\ref{fig:fig1} or time scale over which electronic spin and charge currents, or magnonic spin current, can be computed. Since both electrons and magnons have intrinsic angular momentum, their translational flow leads to a flux of spin angular momentum as spin current.

On the other hand, experiments~\cite{Zilberman1995,Woo2017,Demidov2007,Evelt2017} exciting dipole or exchange dominated SWs   are commonly interpreted using classical micromagnetics~\cite{Kim2010} or atomistic spin dynamics~\cite{Evans2014} simulations (the latter is akin to the former but with atomistic discretization). They describe SWs using trajectories of classical vectors $\mathbf{M}_i(t)$ of fixed (unit) length, pointing along the direction of localized magnetic moments, which precesses around an easy axis with frequency $\omega$ and precession cone angle $\theta$, as illustrated in Fig.~\ref{fig:fig1}. The cone angle has been measured~\cite{Demidov2007} as $\theta \lesssim 10^\circ$. 

In this study,  we employ recently developed multiscale and nonperturbative (i.e., numerically exact)  time-dependent-quantum-transport/classical-atomistic-spin-dynamics  formalism~\cite{Petrovic2018,Bajpai2019a,Petrovic2019} to the problem of electron-SW interaction. This makes possible treating large number of time-dependent localized spins in experimentally relevant noncollinear configurations and over technologically relevant time scales \mbox{$\sim 1$ ns}. The formalism combines time-dependent nonequilibrium Green function (TDNEGF)~\cite{Stefanucci2013,Gaury2014} description of electrons out of equilibrium in open quantum systems, such as those illustrated in Fig.~\ref{fig:fig1}, with the Landau-Lifshitz-Gilbert (LLG) equation describing classical dynamics of localized magnetic moments. The classical treatment of localized magnetic moments is justified~\cite{Wieser2015} in the limit of large localized spins $S \rightarrow \infty$ and $\hbar \rightarrow 0$ (while $S \times \hbar  \rightarrow 1$), as well as in the  absence of entanglement~\cite{Mondal2019} between quantum state of localized spins which is expected to be satisfied at room temperature. 

The paper is organized as follows. In Sec.~\ref{sec:methods} we introduce SW solution and its coupling to quantum Hamiltonian of electrons and TDNEGF calculations. Since explanation of electron-SW scattering for dc injected electronic current [Sec.~\ref{sec:dc}] or leviton current pulse [Sec.~\ref{sec:leviton}] crucially relies on understanding of how SW pumps spin and charge currents in the absence of any bias voltage, we carefully analyze the origin of such pumping in Sec.~\ref{sec:spin_pumping} and Sec.~\ref{sec:charge_pumping}, respectively. This includes computation of nonadiabaticity angle between nonequilibrium electronic spin density and localized magnetic moments in Sec.~\ref{sec:charge_pumping} which visualizes time-retardation effects.   We conclude in Sec.~\ref{sec:conclusion}.

\section{Models and Methods}\label{sec:methods}

Since we assume that a single coherent SW has been excited externally, such as by microwave current flowing through narrow antennas~\cite{Evelt2017}, we fix dynamics of localized magnetic moment $\mathbf{M}_i(t)$ at site $i$ of a one-dimensional (1D) lattice to be 
the SW solution~\cite{Kim2010,Evans2014} of the LLG equation (for simplicity without damping): 
\begin{subequations}\label{eq:sw}
	\begin{eqnarray}
	\mathrm{M}_i^x(t) &=& \sin \theta \ \mathrm{cos}\left( k x_i + \omega t\right),\\
	\mathrm{M}_i^y(t) &=& \sin \theta \ \mathrm{sin}\left( k x_i + \omega t\right),\\
	\mathrm{M}_i^z(t) &=& \cos \theta.
	\end{eqnarray}
\end{subequations}
Due to 1D geometry, the wavevector is just a number $k = 2\pi/[a(N-1)]=2\pi/\lambda$, while the discrete coordinate is $x_i = (i-1)a$ and $N$ is the total number of localized magnetic moments. Note that the {\em uniform mode}--which describes all magnetic moments precessing in-phase in magnetic materials driven by microwaves under the ferromagnetic resonance conditions~\cite{Tserkovnyak2005}---is obtained by setting $k=0$. Even though we employ 1D geometry  as a toy model of a realistic three-dimensional FM layer, such 1D geometries  can be realized experimentally, such as by using an artificial chain of ferromagnetically coupled Fe atoms whose SWs are excited and detected using atom-resolved inelastic tunneling spectroscopy in a scanning tunneling microscope~\cite{Spinelli2014}. We note that solution in Eq.~\eqref{eq:sw} also appears in classical micromagnetics~\cite{Kim2010}, but there $\mathbf{M}_i$ represents magnetization of a small volume of space, typically (2--10 nm)$^3$, rather than of individual atoms~\cite{Evans2014} that we must assume in order to couple classical dynamics of $\mathbf{M}_i(t)$ to time-dependent quantum transport calculations where electrons hop from atom to atom.

The FM nanowire hosting such SW is an active region of devices in Fig.~\ref{fig:fig1}, which is attached to two normal metal (NM) semi-infinite leads  terminating into the  macroscopic reservoirs. We use three different two-terminal geometries depicted Fig.~\ref{fig:fig1}: (a) no bias voltage $V_b$ is applied between the left (L) and right (R) reservoirs kept at the same chemical potential $\mu_\mathrm{L}=\mu_\mathrm{R}$; (b) small dc bias voltage, \mbox{$eV_b=\mu_\mathrm{L} - \mu_\mathrm{R}=0.01$ eV}, is applied between the reservoirs to inject unpolarized charge current into the active region where electrons are spin-polarized by three static localized magnetic moments [red arrows in Fig.~\ref{fig:fig1}(b)] pointing along the $z$-axis; and (c) the Lorentzian voltage pulse~\cite{Ivanov1997,Keeling2006} applied to the left NM lead injects a leviton current pulse $I_\mathrm{L}(t)$ carrying charge $\int\!\! dt\, I_\mathrm{L}(t) = 2e$, which is then spin-polarized by the same three static localized magnetic moments as in (b). 

The quantum Hamiltonian of electrons within the FM nanowire is chosen as 1D tight-binding model
\begin{eqnarray}\label{eq:hamil}
\hat{H}(t) = - \sum_{\braket{ij}} \gamma_{ij} \hat{c}_{i}^{\dagger} \hat{c}_{j} - J_{sd}\sum_i\hat{c}_i^\dagger {\bm \sigma} \cdot \mathbf{M}_i(t) \hat{c}_i,
\end{eqnarray}
with an additional $sd$ exchange interaction of strength \mbox{$J_{sd} = 0.5$ eV}~\cite{Cooper1967} between the spins of the conduction electrons, described by the vector of the Pauli matrices \mbox{${\bm  \sigma} = (\hat{\sigma}_x,\hat{\sigma}_y,\hat{\sigma}_z)$}, and $\mathbf{M}_i(t)$ from Eq.~\eqref{eq:sw}. Here  \mbox{$\hat{c}_i^\dagger=(\hat{c}_{i\uparrow}^\dagger \  \ \hat{c}_{i\downarrow}^\dagger)$} is a row vector containing operators $\hat{c}_{i\sigma}^\dagger$ which create an electron with spin $\sigma=\uparrow,\downarrow$ at site $i$; $\hat{c}_i$ is a column vector containing the corresponding annihilation operators; and \mbox{$\gamma_{ij} = 1$ eV} is the nearest-neighbor hopping. The NM leads are described by the same Hamiltonian as in Eq.~\eqref{eq:hamil} but  with $J_{sd} \equiv 0$. 

The fundamental quantity of nonequilibrium quantum statistical mechanics is the density matrix. The time-dependent one-particle density matrix  can be expressed~\cite{Gaury2014},  ${\bm \rho_{\rm neq}}(t) = \mathbf{G}^<(t,t)/i$, in terms of the lesser GF of TDNEGF formalism  defined by \mbox{$G^{<,\sigma\sigma'}_{ii'}(t,t')=i \langle \hat{c}^\dagger_{i'\sigma'}(t') \hat{c}_{i\sigma}(t)\rangle$} where  
$\langle \ldots \rangle$ is the nonequilibrium statistical average~\cite{Stefanucci2013}. We solve a matrix integro-differential equation~\cite{Popescu2016}  
\begin{equation}
i\hbar\frac{d {\bm \rho}_{\rm neq}}{dt} = [\mathbf{H}(t),{\bm \rho}_{\rm neq}] + i \sum_{ p= L,R} [{\bm \Pi}_p(t) + {\bm \Pi}_p^{\dagger}(t)],
\end{equation}
for the time evolution of  ${\bm \rho_{\rm neq}}(t)$ where $\mathbf{H}(t)$ is the matrix representation of Hamiltonian in Eq.~\eqref{eq:hamil}. This can be viewed as an exact master equation for the reduced density matrix of an open finite-size quantum system attached to macroscopic Fermi liquid reservoirs via semi-infinite NM leads.  The leads ensure continuous energy spectrum of the system and, thereby, dissipation. The ${\bm \Pi}_p(t)$ matrices
\begin{equation}\label{eq:current}
{\bm \Pi}_p(t) = \int_{t_0}^t \!\! dt_2\, [\mathbf{G}^>(t,t_2){\bm \Sigma}_p^<(t_2,t) 
- \mathbf{G}^<(t,t_2){\bm \Sigma}_p^>(t_2,t) ],
\end{equation} 
are expressed in terms of the lesser and greater GF and the corresponding self-energies ${\bm \Sigma}_p^{>,<}(t,t')$~\cite{Popescu2016}. They yield  directly time-dependent total charge, \mbox{$I_p(t) = \frac{e}{\hbar} \mathrm{Tr}\, [{\bm \Pi}_p(t)]$}, and spin, \mbox{$I_p^{S_{\alpha}}(t) = \frac{e}{\hbar} \mathrm{Tr}\, [\hat{\sigma}_{\alpha}{\bm \Pi}_p(t)]$}, currents flowing into the lead $p = \mathrm{L},\mathrm{R}$. Local currents~\cite{Bajpai2019}, or any other local quantity within the active region, are obtained by tracing  the corresponding operator with ${\bm \rho_{\rm neq}}(t)$. We use the same units for charge and spin currents, defined as \mbox{$I_p = I_p^{\uparrow} + I_p^{\downarrow}$} and \mbox{$I_p^{S_{\alpha}} = I_p^{\uparrow} - I_p^{\downarrow}$}, in terms of spin-resolved charge currents $I_p^{\sigma}$. In our convention, {\em positive} current in NM lead $p$ means charge or spin is flowing {\em out} of that lead.

\begin{figure}
	\centering 
	\includegraphics[width= \linewidth]{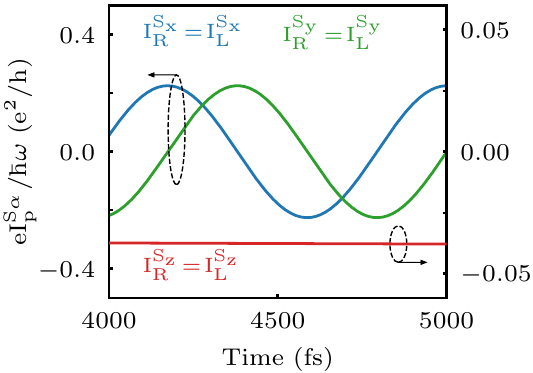}
	\caption{Time-dependence of electronic spin currents \mbox{$I_\mathrm{L}^{S_{\alpha}}(t)=I_\mathrm{R}^{S_{\alpha}}(t)$}  pumped symmetrically~\cite{Tserkovnyak2005,Chen2009} into the L and R NM leads of setup in Fig.~\ref{fig:fig1}(a) whose localized magnetic moments precess as a uniform mode with $k=0$ in Eq.~\eqref{eq:sw}.  The Fermi energy is chosen as \mbox{$E_F=-1.6$ eV}, the frequency of precession is \mbox{$\hbar \omega= 0.005$ eV}, the total number of localized magnetic moments is $N=10$ and dc bias voltage is {\em absent} $V_b \equiv 0$.}
	\label{fig:fig2}
\end{figure}

\section{Results}\label{sec:results}

\subsection{Spin-wave-driven chiral spin pumping} \label{sec:spin_pumping}

As a warm-up, we first consider standard~\cite{Tserkovnyak2005,Chen2009,Mahfouzi2012,Dolui2019} spin pumping by the uniform mode, with $k=0$ in Eq.~\eqref{eq:sw} and no dc bias voltage applied, which will serve as a reference point for subsequent discussion. In this case, identical {\em pure} (i.e., not accompanied by any charge current) spin currents $I_\mathrm{L}^{S_\alpha}(t)=I_\mathrm{R}^{S_\alpha}(t)$ are pumped into both leads, as shown in Fig.~\ref{fig:fig2}. Their $I_\mathrm{L}^{S_z}=I_\mathrm{R}^{S_z}$ components are time independent, and their negative sign shows that they flow {\em into} the NM leads, as obtained also in the scattering theory~\cite{Tserkovnyak2005}, rotating frame approach~\cite{Chen2009} or Floquet-NEGF theory~\cite{Mahfouzi2012,Dolui2019}.

On the other hand, the excited SW in the setup of Fig.~\ref{fig:fig1}(a)  pumps {\em both} charge and spin currents into the NM leads in the absence of any dc bias voltage. Their time dependences, $I_p^{S_{\alpha}}(t)$ and $I_p(t)$, are shown in Fig.~\ref{fig:fig2} after transient currents have died out. Furthermore, in contrast to pumping by the uniform mode, we find $|I_\mathrm{L}^{S_z}| > |I_\mathrm{R}^{S_z}|$. This is due to the spin current carried by the SW itself~\cite{Sandweg2011}. That is, spin current carried by the SW  must be ``transmuted''~\cite{Bauer2011}  into electronic spin current at the FM-wire/NM-left-lead interface~\cite{Sandweg2011,Woo2017,Petrovic2019} because no localized magnetic moments exist in the NM lead to support transport of angular momentum via their dynamics. This current is then added or subtracted  to symmetrically pumped spin currents into the left or right NM leads, respectively. This interpretation is supported by the fact that changing the sign of $k$ in Eq.~\eqref{eq:sw} leads to a reversed situation, $|I_\mathrm{L}^{S_z}| < |I_\mathrm{R}^{S_z}|$.

\begin{figure}
	\centering 
	\includegraphics[width= \linewidth]{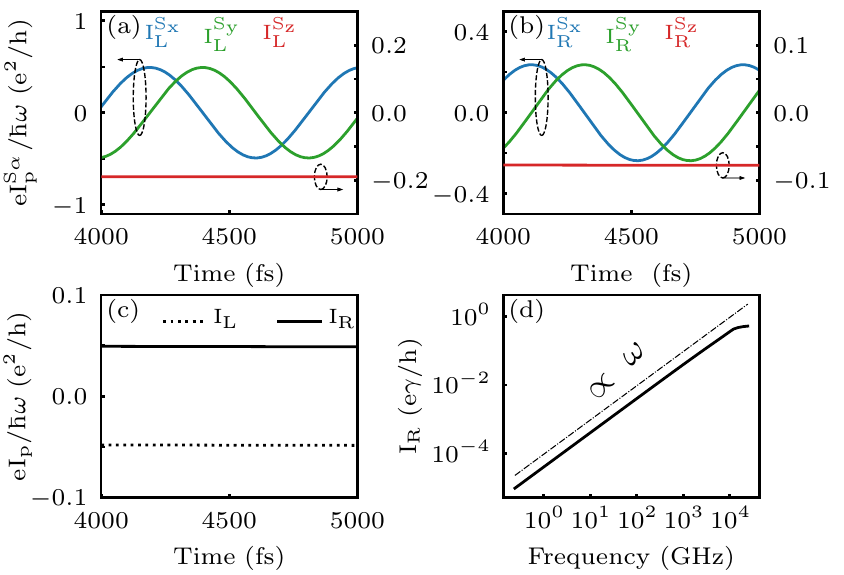}
	\caption{Time-dependence of electronic spin currents pumped into the (a) left and (b) right NM leads of setup in Fig.~\ref{fig:fig1}(a) whose localized magnetic moments precess as coherent SW mode with $k \neq 0$ in Eq.~\eqref{eq:sw}. (c) The SW with its nonuniform precessing magnetic moments also pumps dc charge current $I_\mathrm{L} = - I_\mathrm{R}$ into the NM leads, whose dependence on frequency (solid line) in panel (d) is linear $\propto \omega$ (dash-dot line).  The Fermi energy is chosen as \mbox{$E_F=-1.6$ eV}, the frequency of SW is \mbox{$\hbar \omega= 0.005$ eV}, the total number of localized magnetic moments is $N=10$ and dc bias voltage is {\em absent}, $V_b \equiv 0$.}
	\label{fig:fig3}
\end{figure}

\subsection{Spin-wave-driven chiral charge pumping} \label{sec:charge_pumping}

The charge pumping in spintronic devices with excited coherent SWs was observed experimentally in compressively strained (Ga,Mn)As bar~\cite{Ciccarelli2014}, as well as in YIG/graphene~\cite{Evelt2017}. In the latter case,  SW is excited within insulating YIG while pumped current flows through metallic graphene where localized magnetic moments are induced by proximity exchange coupling~\cite{Hallal2017}. However, both of these experimental setups require SOC, unlike our setup in Fig.~\ref{fig:fig1}(a) where SOC is {\em absent}. 

In the {\em adiabatic} limit~\cite{Stahl2017}, the conduction electron spin at site $i$, $\langle \hat{\mathbf{s}}_i \rangle_t= \frac{\hbar}{2} \mathrm{Tr} \, [{\bm \rho}_\mathrm{neq}(t) |i\rangle \langle i| \otimes {\bm \sigma}]$, follows strictly the direction of localized magnetic moment $\mathbf{M}_i(t)$ at the same site. In this limit, the charge pumping by  time-dependent noncoplanar and noncollinear magnetic texture, described by local magnetization $\mathbf{m}(\mathbf{r},t)$ as a continuous variable, is predicted by the spin motive force (SMF) theory~\cite{Volovik1987,Barnes2007,Duine2008,Tserkovnyak2008a,Zhang2009b}
\begin{equation}\label{eq:smf}
j_\alpha (\mathbf{r}) = C [\partial_\alpha \mathbf{m}(\mathbf{r},t) \times \mathbf{m}(\mathbf{r},t)] \cdot  \partial_t \mathbf{m}(\mathbf{r},t). 
\end{equation}
This formula is rooted in the associated geometrical Berry phase. Here ${\bm \rho}_\mathrm{eq}$ is the equilibrium density matrix;  $C=P G_0 \hbar/2e$;   \mbox{$P=(G^\uparrow - G^\downarrow)/(G^\uparrow + G^\downarrow)$} is the spin polarization of the ferromagnet; and 
$G_0=G^\uparrow + G^\downarrow$ is the total conductivity. We use notation $\partial_t = \partial/\partial t$ and $\partial_\alpha=\partial/\partial \alpha$ for $\alpha \in \{x,y,z \}$. If we plug SW solution for the local magnetization---$m_x(x) = \sin \theta \cos (kx + \omega t)$; $m_y(x) = \sin \theta \sin (kx + \omega t)$; and $m_z(x) = \cos \theta$---into Eq.~\eqref{eq:smf} we obtain {\em zero} pumped charge current, $j_x (x) \equiv 0$. 

It is worth clarifying that if plug SW solution from Eq.~\eqref{eq:sw} into the discretized version of Eq.~\eqref{eq:smf}  
\begin{equation}\label{eq:smfdiscrete}
j_x(i) = \frac{C}{a} \left[\partial_t \mathbf{M}_i \times \mathbf{M}_{i+1}\right] \cdot \mathbf{M}_i,
\end{equation}
we actually obtain a nonzero result, \mbox{$j_x(i) = \frac{C\omega}{a} \sin\theta \sin2 \theta \sin^2 (ka/2)$}. This apparently contradicts $j_x (x) \equiv 0$ obtain from the continuous formula in Eq.~\eqref{eq:smf}. However, in the limit of the lattice spacing going to zero, $a \rightarrow 0$, we have $\lim_{a \to 0}\frac{1}{a} \sin^2(ka/2) = 0$ and, therefore, the same conclusion, $j_{x}(i) \equiv 0$.

\begin{figure}
	\includegraphics[width= \linewidth]{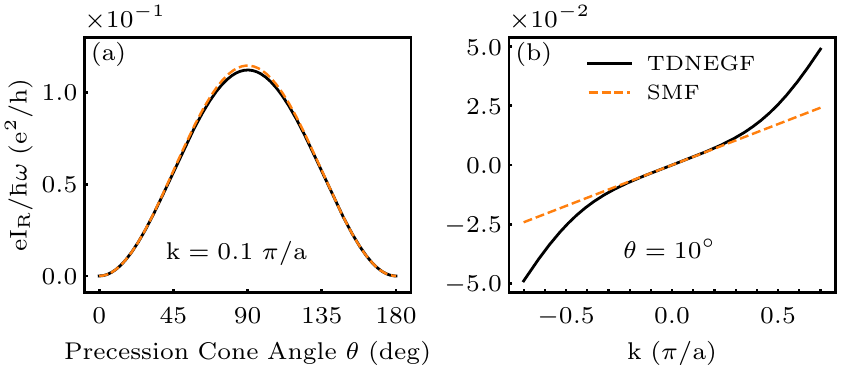}
	\caption{The dependence of charge current from Fig.~\ref{fig:fig3}(c) on: (a) precession cone angle $\theta$; and (b) wavevector $k$ of the SW. The solid lines are obtained from TDNEGF calculations and the dashed line is obtained from modified~\cite{Evelt2017} SMF formula in Eq.~\eqref{eq:smflambdasw}.}
	\label{fig:fig4}
\end{figure}
\begin{figure*}
	\includegraphics[width= \linewidth]{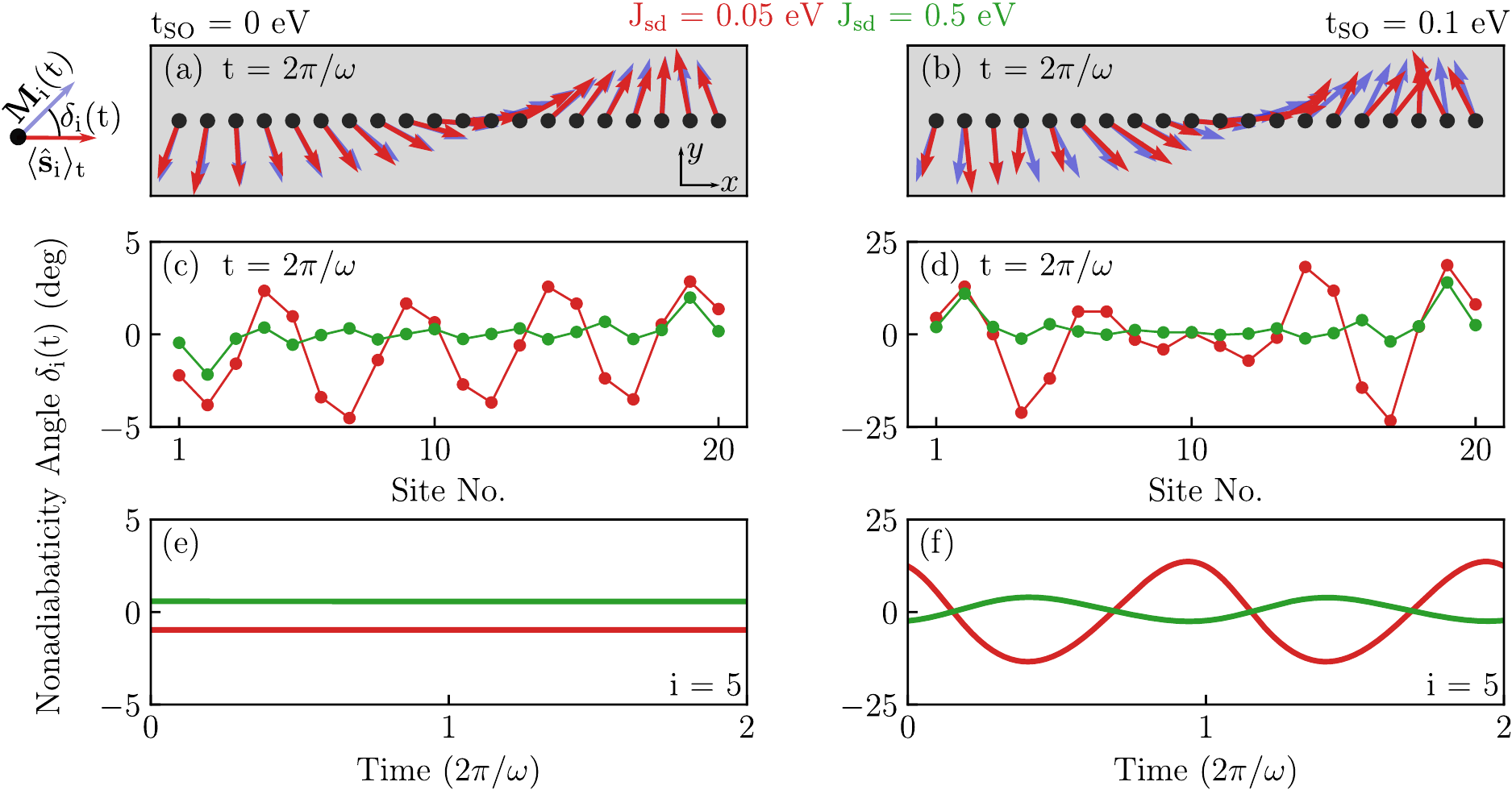}
	\caption{(a),(b) Spatial profile at time $t=2\pi/\omega$ of the in-$xy$-plane component of nonequilibrium electronic spin density vector $\langle \hat{\mathbf{s}}_i \rangle_t$ and localized magnetic moments $\mathbf{M}_i(t)$ across FM nanowire  hosting a SW of wavevector $k \simeq 0.2a$. Inset on the left illustrates the nonadiabaticity angle $\delta_i$ between these two vectors. (c),(d) Spatial profile of nonadiabaticity angle $\delta_i(t)$ at time $t=2\pi/\omega$. (e),(f) Time dependence of $\delta_{i=5}(t)$ at site $i=5$. In panels (a),(c),(e) SOC is absent, while in panels (b),(d),(f) we use Rashba SOC in Eq.~\eqref{eq:rashba} of strength \mbox{$t_\mathrm{SO}=0.1$ eV}. Two different values of $sd$ exchange interaction in the Hamiltonian [Eq.~\eqref{eq:hamil}] are used, \mbox{$J_{sd}=0.05$ eV} (red lines) and \mbox{$J_{sd}=0.5$ eV} (green lines). The Fermi energy is chosen as \mbox{$E_F=-1.6$ eV}, the frequency of SW is \mbox{$\hbar \omega= 0.005$ eV}, the total number of localized magnetic moments is $N=20$ and dc bias voltage is {\em absent}, $V_b \equiv 0$.}
	\label{fig:angle}
\end{figure*}

In Ref.~\cite{Evelt2017}, a modified version of the SMF formula
\begin{equation}\label{eq:smflambda}
j_\alpha (\mathbf{r}) = C [\partial_\alpha \mathbf{m}(\mathbf{r},t) \times \mathbf{m}(\mathbf{r},t) + \beta \partial_\alpha \mathbf{m}(\mathbf{r},t)] \cdot  \partial_t \mathbf{m}(\mathbf{r},t), 
\end{equation}
was employed to explain the experiment. Here adding {\em nonadiabatic} correction of magnitude $\beta$ to purely geometrical first term was justified~\cite{Evelt2017} as the consequence of slight misalignment between electron spin and localized magnetic  moments caused by SOC and thereby induced relaxation of nonequilibrium electron spin density. Thus, using SW  solution in  the discretized version of $\beta$-term in Eq.~\eqref{eq:smflambda} gives
\begin{eqnarray}\label{eq:smflambdasw}
j_x^\beta(i) & = & C\beta \partial_t \mathbf{M}_i \cdot \left( \frac{\mathbf{M}_{i+1} - \mathbf{M}_i}{a}  \right) \nonumber \\
                 & = & C\beta \omega \sin^2\theta \,  \left( \frac{\sin ka}{a} \right) \xrightarrow[a \to 0]{}  C k \omega \sin^2\theta.
\end{eqnarray}
The final result explains experimentally observed~\cite{Evelt2017} chiral nature of pumping where charge current changes sign upon $k \rightarrow -k$. 

We compare this result with the one from TDNEGF calculations in Fig.~\ref{fig:fig4}, where they track each other [Fig.~\ref{fig:fig4}(a)] as a function of cone angle $\theta$, except around $\theta = 90^\circ$; as well as as a function of $k$ [Fig.~\ref{fig:fig4}(a)] within $|k| \lesssim 0.2\pi/a$ interval. Since our TDNEGF calculations are numerically exact, such deviations (for values of $\theta$ and $k$ not commonly found in experiments though~\cite{Woo2017,Demidov2007}) stem from the fact that the SMF formula in Eq.~\eqref{eq:smflambda} contains only the lowest order~\cite{Duine2008} time and spatial derivatives of local magnetization. 

This also demonstrates how TDNEGF calculations automatically include nonadiabatic effects in spin dynamics even when SOC is zero. The origin of $\beta$-term in Eq.~\eqref{eq:smflambda} in the absence of SOC, which is effectively generated by TDNEGF calculations in Fig.~\ref{fig:fig4},  is that direction of nonequilibrium electron spin density at site $i$, $\langle \hat{\mathbf{s}}_i \rangle_t$, is always somewhat behind the `adiabatic direction' set by the classical vector $\mathbf{M}_i(t)$. So, the nonadiabaticity results from the fact that the motion of the classical spin affects the conduction electrons in a {\em retarded way}~\cite{Sayad2015}---it takes a finite time until the local conduction electron spin $\langle \hat{\mathbf{s}}_i \rangle_t$ reacts to the motion of the classical spin. This is visualized in Fig.~\ref{fig:angle}(a),(c) where nonadiabaticity angle $\delta_i(t)$ between $\langle \hat{\mathbf{s}}_i \rangle_t$ and $\mathbf{M}_i(t)$ decreases with increasing $J_{sd}$ (realistic values measured experimentally are \mbox{$J_{sd} \simeq 0.1$ eV}~\cite{Cooper1967}). In the absence of SOC, the angle $\delta_i(t)$ in Fig.~\ref{fig:angle}(e) is time independent. The resulting spin torque, $\propto \langle \hat{\mathbf{s}}_i \rangle_t \times \mathbf{M}_i(t)$, exerted on the classical localized magnetic moment acts then like an additional Gilbert damping which is, generally, described by nonlocal-in-time damping kernel with memory effects~\cite{Bajpai2019a,Sayad2015,Hurst2020}. Note, however, that we fix the dynamics of localized magnetic moments to SW solutions in Eq.~\eqref{eq:sw}, rather than solving TDNEGF and LLG equations self-consistently~\cite{Petrovic2018,Bajpai2019a,Petrovic2019}.

For comparison, in Fig.~\ref{fig:angle}(b),(d) we use additional Rashba SOC term~\cite{Manchon2015} in the quantum Hamiltonian in Eq.~\eqref{eq:hamil}
\begin{equation}\label{eq:rashba}
\hat{H}_\mathrm{SO} =  \sum_{ij} \hat{c}_i^\dagger \mathbf{t}_{ij}\hat{c}_j,
\end{equation}
where $\mathbf{t}_{ij} = - i t_\mathrm{SO} \hat{\sigma}_y$ for $i = j + 1$. The strength of the Rashba SOC is chosen as $t_\mathrm{SO}=0.1$ eV, which generates conventional static Gilbert damping $\alpha_\mathrm{G}=0.01$ via the scattering theory~\cite{Brataas2011} as the typical value~\cite{Taniguchi2015} found in FM nanowires. This additional Rashba SOC leads to increase of $\delta_i$ in Fig.~\ref{fig:angle}(b),(d) when compared to Fig.~\ref{fig:angle}(a),(c), respectively. It also generates periodic time dependence of $\delta_i(t)$ in Fig.~\ref{fig:angle}(f).

\begin{figure}
	\includegraphics[width= \linewidth]{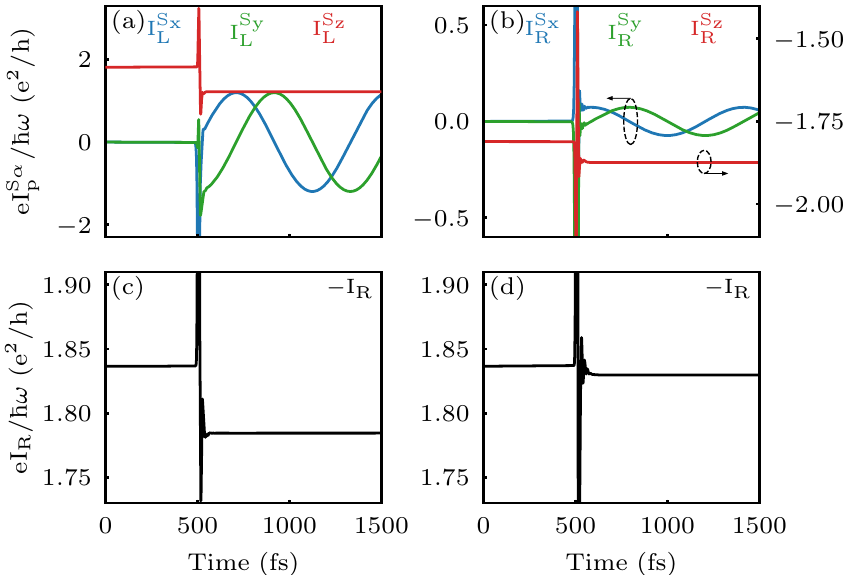}
	\caption{Time-dependence of electronic spin currents in the (a) left and (b) right NM leads of setup in Fig.~\ref{fig:fig1}(b) whose localized magnetic moments start to precess at $t = 500$~fs as a coherent SW with $k \neq 0$ in Eq.~\eqref{eq:sw} in the presence of a flux of  electrons injected into the active region by dc bias voltage \mbox{$eV_b = 0.01$ eV}. The electrons are spin-polarized by three fixed spins (red arrows) in Fig.~\ref{fig:fig1}(b). The corresponding time-dependence of their charge current $I_\mathrm{R}(t)$ is shown in panel (c). Panel (d) plots time dependence of $I_\mathrm{R}(t)$ for ``frozen-in-time''~\cite{Starikov2018} SW where $t=0$ in Eq.~\eqref{eq:sw}. The Fermi energy is chosen as \mbox{$E_F=-1.6$ eV}, the frequency of SW is \mbox{$\hbar \omega= 0.005$ eV}, the total number of localized magnetic moments is $N=10$ and dc bias voltage is \mbox{$V_b = 0.01$ V}.}
	\label{fig:fig5}
\end{figure}

Another way to interpret the origin of charge pumping by SW is to analyze its frequency dependence shown in Fig.~\ref{fig:fig3}(d). This complies with the general theory of ``adiabatic'' quantum pumping~\cite{Moskalets2002,FoaTorres2005,Bajpai2019} since it scales linearly with frequency in the physically relevant frequency range GHz--THz~\cite{Jungfleisch2018}. Note that terminology ``adiabatic'' in this context is not related to spin (unlike the preceding discussion where adiabatic means that flowing electron spin and localized spins are aligned instantaneously~\cite{Stahl2017})---instead it signifies sufficiently slow change of harmonic potential driving the quantum system so that its frequency is $\hbar \omega \ll E_F$ and/or  smaller than relevant relaxation time for electrons.  In contrast, charge pumping by SW in YIG/graphene heterostructure with SOC peaks between 5 and 7 GHz~\cite{Evelt2017}. We recall that such linear scaling is in accord with the key requirement---{\em breaking of left-right symmetry}---for nonzero dc component of quantum charge pumping by a time-dependent potential~\cite{Moskalets2002,FoaTorres2005,Bajpai2019}. This can be achieved by breaking inversion symmetry and/or time-reversal symmetry.  In the ``adiabatic''  regime, quantum charge pumping requires both inversion and time-reversal symmetries to be broken dynamically, such as by two spatially separated potentials oscillating out-of-phase~\cite{Moskalets2002}, which leads to $\bar{I}_p(t) \propto \omega$  at low frequencies ($\bar{A}$ is the average of quantity $A$ over one period). In the case of SW, it is the wave-like pattern of precessing localized magnetic moments which dynamically breaks the left-right symmetry in Fig.~\ref{fig:fig1}(a), with respect to vertical plane positioned between moments localized at sites $i=N/2$ and $i=N/2+1$. In contrast, in the ``nonadiabatic'' regime only one of those two symmetries needs to be broken and this does not have to occur dynamically. The dc component of the pumped current in the ``nonadiabatic`` regime is~\cite{FoaTorres2005} $\bar{I}_p(t) \propto \omega^2$ at low frequencies as obtained in, e.g.,  the case of charge pumping by uniform mode across potential barrier that breaks the left-right symmetry of the device statically~\cite{Chen2009}. 

\begin{figure}
	\centering
	\includegraphics[width= \linewidth]{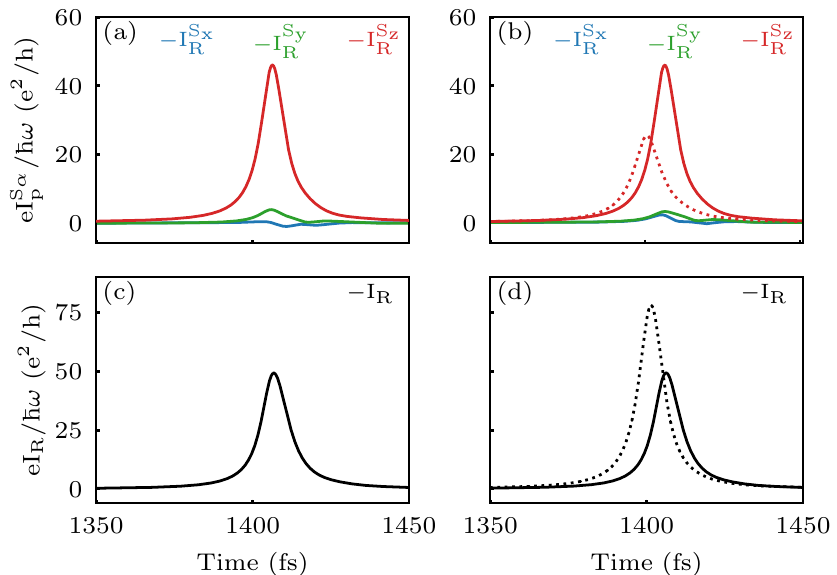}
	\caption{Time-dependence of electronic spin currents $I_\mathrm{R}^{S_{\alpha}}(t)$  in the right NM lead after unpolarized leviton is injected  by the Lorentzian voltage pulse~\cite{Ivanov1997,Keeling2006} into the active region hosting: (a) three static localized magnetic moments (red arrows) pointing  along the $z$-axis, acting as spin-polarizer, followed by SW excited at \mbox{$t=500$ fs}, as illustrated in Fig.~\ref{fig:fig1}(c); (b) three static localized magnetic moments  and a ``frozen-in-time'' SW excited at \mbox{$t=500$ fs}. Panels (c) and (d) show time dependence of the charge current  $I_\mathrm{R}(t)$ corresponding to (a) and (b), respectively. In addition, panels (b) and (d) plot (dotted lines) time dependence of $I_\mathrm{R}^{S_{\alpha}}(t)$ and $I_\mathrm{R}(t)$, respectively, for the active region containing only the three static localized magnetic moments pointing  along the $z$-axis. The Fermi energy is chosen as \mbox{$E_F=-1.6$ eV}, the frequency of SW is \mbox{$\hbar \omega= 0.005$ eV} and the total number of localized magnetic moments is $N=10$.}
	\label{fig:fig6}
\end{figure}

\subsection{Electron/spin-wave scattering for injected dc spin-polarized charge current}\label{sec:dc}

For the setup in Fig.~\ref{fig:fig1}(b), we first establish (after some transient period not shown explicitly) steady  charge current $I_\mathrm{R}$ [flat line in Fig.~\ref{fig:fig5}(c) for $t<500$ fs] of electrons injected by dc bias voltage into $3+10$ static localized magnetic moments oriented along the $z$-axis. The initially unpolarized current becomes spin-polarized due to static moments, as characterized by steady spin current $I_\mathrm{R}^{S_z} \neq 0$ [flat line in Fig.~\ref{fig:fig5}(b) for $t<500$ fs] and the corresponding spin-polarization $P_z = |I_\mathrm{R}^{S_z}|/|I_\mathrm{R}| \approx 50$\%. Then at \mbox{$t=500$ fs} we {\em suddenly} excite SW composed of  $10$ precessing localized magnetic moments in Fig.~\ref{fig:fig1}(b). This induces transient currents around that instant which help us to visualize the boundary between the time interval without and with SW being present. Within the time interval \mbox{$t>500$ fs} where SW is present, new time-dependent spin currents $I_p^{S_x}(t)$ and $I_p^{S_y}(t)$ emerge [Fig.~\ref{fig:fig5}(a),(b)] due to spin pumping by SW demonstrated in Fig.~\ref{fig:fig3}(a),(b).

Concurrently, dc spin currents $I_\mathrm{L}^{S_z}$ [Fig.~\ref{fig:fig5}(a)] and $I_\mathrm{R}^{S_z}$ [Fig.~\ref{fig:fig5}(b)], as well as dc charge current $I_\mathrm{R}$ [Fig.~\ref{fig:fig5}(c)], are reduced compared to their values prior to SW excitation. This reduction is mostly due to charge and spin currents pumped in the direction right-NM-lead$\rightarrow$left-NM-lead in Fig.~\ref{fig:fig3}, which is opposite to the flow of originally injected charge and spin currents by dc bias voltage. Thus, outflowing spin and charge currents in the right NM lead can also be enhanced if we invert the sign of $k$ in Eq.~\eqref{eq:sw} and, therefore, the direction of SW propagation. Another reason for the reduction is backscattering of electrons by time-dependent potential generated by SW, whose magnitude for charge current shown in Fig.~\ref{fig:fig5}(c) we estimate using \mbox{$[I_\mathrm{R}(t<500 \ \mathrm{fs}) - I_\mathrm{R}(t>500 \ \mathrm{fs}) + I_\mathrm{R}^\mathrm{SW}]/I_\mathrm{R}(t<500 \ \mathrm{fs})$} to be less than $1$\%. Here $I_\mathrm{R}^\mathrm{SW}$ denotes charge current pumped by SW [Fig.~\ref{fig:fig3}(c)] in the absence of dc bias voltage.

Recent {\em time-independent} quantum transport calculations~\cite{Starikov2018} of the resistance of FM have include ``frozen magnons'' as correlated spin disorder where localized spins are tilted away from the easy axis in accord with thermal population of magnon modes. To understand time-dependent effects missed in such  calculations, we freeze localized magnetic moments by setting $t=0$ in Eq.~\eqref{eq:sw}. The scattering from such ``frozen-in-time'' SW leads to much smaller current reduction in Fig.~\ref{fig:fig5}(d).  

\subsection{Electron/spin-wave scattering for injected spin-polarized charge current leviton pulse}\label{sec:leviton}

In order to simulate single-electron/single-SW scattering, we inject pulsed current into the active region using the Lorentzian voltage pulse~\cite{Popescu2016}, \mbox{$V_\mathrm{L}(t) = 2\hbar \tau/[(t-t_0)^2 + \tau^2]$}, where the  pulse duration is $\tau = 7.5 \hbar/\gamma$. As confirmed experimentally~\cite{Dubois2013}, such special pulse profile with  $\frac{e}{\hbar} \int \!\! dt \, V_\mathrm{L}(t) = 2 \pi n$~\cite{Moskalets2016} drives the Fermi sea in the left reservoir to ensure~\cite{Ivanov1997,Keeling2006} excitation of an integer number $n$ of purely electronic states above the sea. They appear without  spurious electron-hole pairs and exhibit minimal~\cite{Gaury2016} nonequilibrium noise in charge transfer across the active region. We use $n=2$, so that injected unpolarized charge current pulse, called leviton~\cite{Dubois2013}, carries charge $\int \!\! dt \,  I_\mathrm{L}(t)=2e$. This can be viewed as minimalistic unpolarized current composed of one spin-$\uparrow$ and one spin-$\downarrow$ electron flowing together. Upon spin-polarization by three static localized magnetic moments in Fig.~\ref{fig:fig1}(c), the leviton interacts with SW  excited suddenly at  \mbox{$t = 500$ fs}. After such interaction, leviton outflows into the right NM lead where its spin and charge currents are plotted in Figs.~\ref{fig:fig6}(a) and ~\ref{fig:fig6}(c), respectively. For comparison, Figs.~\ref{fig:fig6}(b) and ~\ref{fig:fig6}(d) plot spin and charge currents in the right NM lead, respectively, for a leviton interacting with ``frozen-in-time'' SW. In addition, Figs.~\ref{fig:fig6}(b) and ~\ref{fig:fig6}(d) also include (dotted lines) spin and charge current of  leviton  outflowing into the right NM lead when SW in Fig.~\ref{fig:fig1}(c) is removed from the active region. The integrals for outflowing spin-polarized leviton after scattering from SW in Figs.~\ref{fig:fig6}(a) and ~\ref{fig:fig6}(c) are $\int \!\! dt \,  I_\mathrm{R}^{S_z}(t) = 0.939e$ and $\int \!\! dt \,  I_\mathrm{R}(t)  = 0.822e$, respectively. They can be compared to $\int \!\! dt \,  I_\mathrm{R}^{S_z}(t) = 0.944e$ and $\int \!\! dt \,  I_\mathrm{R}(t) = 0.827e$ in Figs.~\ref{fig:fig6}(b) and ~\ref{fig:fig6}(d), respectively. Note that the the ratio of integrals of two dotted curves in Figs.~\ref{fig:fig6}(b) and ~\ref{fig:fig6}(d) is $P_z = | \int \!\! dt \,  I_\mathrm{R}^{S_z}(t)|/ |\int \!\! dt \,  I_\mathrm{R}(t)| \approx 40$\% which can be considered as the   spin-polarization of leviton after passing through three static localized magnetic moments in Fig.~\ref{fig:fig1}(c).

\section{Conclusions}\label{sec:conclusion}
In conclusion, using time-dependent-quantum-transport/classical-atomistic-spin-dynamics multiscale framework~\cite{Petrovic2018,Bajpai2019a,Petrovic2019} we predict that SW coherently excited within a metallic ferromagnet will pump  {\em chiral} electronic charge and spin currents into the attached normal metal leads. The chirality of pumped currents means that their direction is tied to the direction of SW propagation, changing upon reversal of the SW wavevector. The pumped currents scale linearly with the frequency of the SW in experimentally relevant GHz--THz range. In contrast, recent experiments on ``magnonic charge pumping''~\cite{Ciccarelli2014,Evelt2017} were interpreted by requiring nonzero SOC to introduce misalignment between electron spin and localized magnetic moments, thereby adding  nonadiabatic contribution~\cite{Evelt2017} to the spin motive force formula~\cite{Volovik1987,Barnes2007,Duine2008,Tserkovnyak2008a,Zhang2009b}. This formula describes charge pumping by time-dependent noncoplanar and noncollinear magnetic textures. Although SW is an example of such texture, standard purely adiabatic (i.e., for electron spin and localized spins aligned instantaneously) spin motive force formula [Eq.~\eqref{eq:smf}] valid in the absence of SOC predicts zero pumped charge current [Sec.~\ref{sec:charge_pumping}]. Thus our prediction reveals the importance of time-retardation effects~\cite{Bajpai2019a,Sayad2015}, where conduction electron spin is always somewhat behind the `adiabatic direction' set by the classical localized spins. The retardation is visualized [Fig.~\ref{fig:angle}] by plotting the nonadiabaticity angle between conduction electron spin and classical localized magnetic moment, both in the absence and presence of SOC. When dc spin-polarized charge current is injected into the ferromagnet, electrons interact with SW in such a way that the outflowing charge and spin current are changed both by the scattering off time-dependent potential generated by the SW and superposition with the currents pumped by the SW itself. Using Lorentzian voltage pulse to excite leviton out of the Fermi sea, which carries one electron charge with no accompanying electron-hole pairs and behaves as soliton-like quasiparticle, we also show how a {\em single} electron scatters from a {\em single} SW. 

\begin{acknowledgments}
This research was supported in part by the US National Science Foundation (NSF) under Grant No. ECCS-1922689. It was finalized during {\em Spin and Heat Transport in Quantum and Topological Materials} program at KITP Santa Barbara, which is supported under NSF Grant No. PHY-1748958.
\end{acknowledgments}

%********************references***************************

%
%\bibliographystyle{ieeetran}
%\bibliography{mybib.bib}
%

\end{document}